\documentclass[pr,twocolumn,showpacs,longbibliography]{revtex4}
\usepackage{bm}
\usepackage{graphicx}
\usepackage{amssymb}
\usepackage{amsmath}
\usepackage{eufrak}
\usepackage{color}
\usepackage[utf8]{inputenc}
\usepackage{pifont}
\usepackage{ulem}
\usepackage[unicode=true,colorlinks=true,citecolor=blue,urlcolor=blue]{hyperref}

\renewcommand{\Im}{\mathop{\rm Im}}

\newcommand{\e}{\mathrm{e}}

\renewcommand{\i}{{\rm i}}
\renewcommand{\d}{\mathrm d}
\renewcommand{\emph}{\textit}

\newcommand{\ttr}{{t_{\rm tr}}}
\newcommand{\odr}{{\Omega_{\rm dr}}}

\newcommand{\enquote}{}

\newcommand{\nix}[1]{}

\begin{document}

\title{
Spin dynamics and fluctuations in the streaming regime}
\author{D.\,S. Smirnov and L.\,E.~Golub}
\affiliation{Ioffe Institute, 194021 St.~Petersburg, Russia}

\begin{abstract}
Spin dynamics of two-dimensional electrons in moderate in-plane electric fields is studied theoretically. The streaming regime is considered, where each electron accelerates until reaching the optical phonon energy, then it emits an optical phonon, and a new period of acceleration starts.
Spin-orbit interaction and elastic scattering result in anisotropic relaxation of electron spin polarization. 
The overall spin dynamics is described by a superposition of spin modes in the system. 
The relaxation time of the most long-living mode depends quasi-periodically on the inverse electric field.
The spin modes can be conveniently revealed by means of spin noise spectroscopy.
It is demonstrated that the spectrum of spin fluctuations consists of peaks with the low-frequency peak much narrower than satellite ones, and the widths of the peaks are determined by the decay times of the modes.
\end{abstract}

\pacs{72.25.Hg, 72.25.Rb, 72.70.+m, 73.63.Hs, 78.47.db}

\maketitle{}

\section{Introduction}
\label{intro}

\begin{figure}[t]
\includegraphics[width=\linewidth]{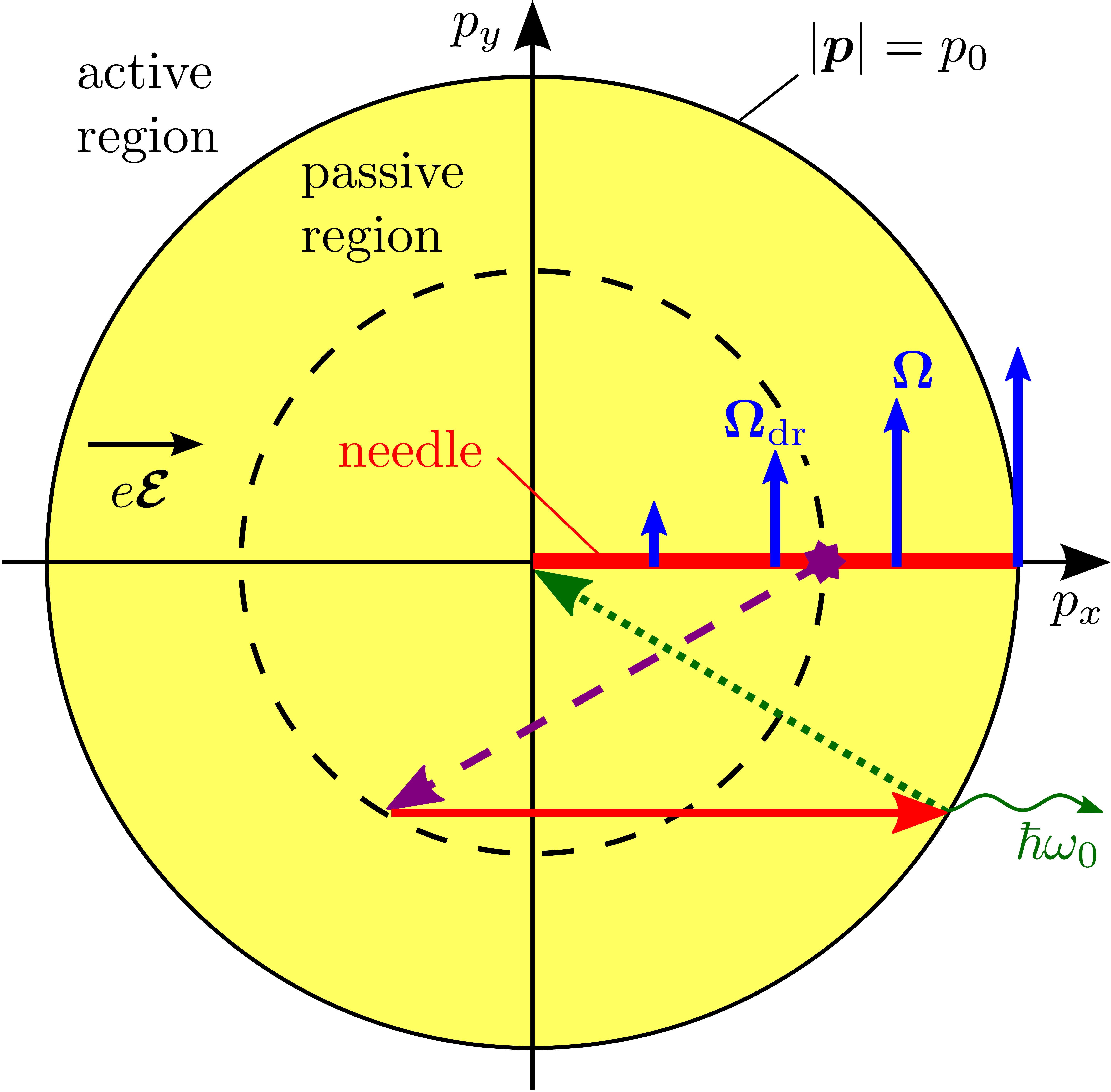}
\caption{Illustration of the streaming regime in the momentum space. During a ballistic motion inside the needle (red solid line), an electron spin rotates with the frequency $\bm \Omega(\bm p)$
(blue arrows). The electron can be elastically scattered by an impurity (star and dashed arrow), then it reaches the active area (red arrow), instantaneously emits an optical phonon (wavy arrow) and returns to $\bm p=0$ (dotted arrow).}
\label{fig:illustration}
\end{figure}

A moderately strong electric field applied to a semiconductor system with low carrier concentration can provide the \textit{streaming} regime of electron transport~\cite{Andronov}. In this regime, each free charge carrier accelerates quasiballistically in the passive region of the momentum space, where its energy is smaller than the optical-phonon energy $\hbar\omega_0$. 
As soon as the carrier energy amounts to $\hbar\omega_0$, it
emits an optical phonon and  scatters to a state with a small energy. This is the end of the period, after which the next cycle of acceleration starts. The electron momentum changes periodically from zero to a value $p_0$ corresponding to the electron energy equal to $\hbar\omega_0$, Fig.~\ref{fig:illustration}. The period of oscillations is the electron travelling time from $p=0$ to $p=p_0$:
\begin{equation}
	\ttr = p_0/|e\cal{E}|, 
\end{equation}
where $e<0$ is the electron charge and $\cal{E}$ is the electric field strength. The electron distribution in the momentum space is strongly anisotropic, {i.e. its} length in the field direction, $p_0$, {is} much larger than {its} width. {The corresponding region of the momentum space is called a \textit{needle} and is shown by the red solid line in Fig.~\ref{fig:illustration}.}

This regime of carrier transport has been studied in detail in three-dimensional systems. 
A semiconductor laser in \textit{p}-Ge has been realized based on the streaming effect~\cite{Andronov_Gornik}.
Many interesting features can be revealed by studying the electric current fluctuations in the streaming regime~\cite{Bareikis,Kogan}. 
Recent theoretical investigations of streaming in two-dimensional systems demonstrate collective wave-like excitations of the electrons with multi-branch spectra and considerable spatial dispersion~\cite{Korot_2012}.
Ballistic transport with dominant optical phonon scattering has been realized in graphene~\cite{Barreiro2009}, and a number of interesting theoretical proposals have been made for graphene~\cite{Fang2011,Sekwao2011,Sekwao2013}.

The inclusion of electron spin degree of freedom into the two-dimensional streaming-regime kinetics gives rise to rich spin-related phenomena. 
Due to electron drift in the electric field and linear in momentum Rashba and Dresselhaus spin-orbit interactions, electron spin precess with an average frequency $\odr$. 
Electrically induced spin beats and long spin relaxation times as well as high degree of the current induced spin polarization have been predicted for such a system~\cite{Golub2013,Golub2014}.
However the comprehensive study of spin dynamics in the regime of substantial spin rotations in each acceleration period ($\odr\ttr\sim1$) has not been made yet.
As we show, the spin dynamics is not reduced to a simple exponential relaxation, and the electron spin polarization at resonant conditions persists despite of multiple elastic scatterings.

Estimations show that the streaming regime in GaAs based heterostructures can be realized in reasonable fields ${\cal E} \sim 1$~kV/cm, where the parameter $\odr\ttr$ is of the order of unity. This shows a possibility of experimental investigations of spin-dependent phenomena in the streaming regime.

The spin-dependent streaming kinetics has a few temporal ranges: {$\ttr$, ${\rm \Omega}_{\rm dr}^{-1}$}, the elastic scattering time $\tau_p$, and the spin relaxation time {$\tau_s$}. 
Therefore it is natural to study spin dynamics in the frequency domain. This can be done by means of spin noise spectroscopy being a modern and very efficient tool for investigation of spin properties in various systems~\cite{Zapasskii:13,Oestreich-review,Zapasskii13}. 
This method is based on the measurement of fluctuating spin signals in the ensemble of unpolarized carriers
and allows simultaneous resolving different time ranges~\cite{noise-excitons}.
The spin noise of free two-dimensional carriers in electric fields has been considered in nearly equilibrium conditions where the electric field is weak~\cite{Sinitsyn,Slipko2013}. 
Here we investigate spin fluctuations in heterostructures in the streaming regime which represents the opposite situation of strongly non-equilibrium electron gas. We demonstrate that the spectrum of spin fluctuations in this case consists of a series of peaks with different widths.

The paper is organized as follows. In Sec.~\ref{theory}, {the general} theory of spin dynamics is developed for the two-dimensional streaming regime with account for spin-orbit interaction and elastic scattering. In Section~\ref{results}, we consider the spin dynamics 
in the presence of either longitudinal or transverse effective field.
In Sec.~\ref{noise}, the spin noise in the streaming regime is investigated. Concluding remarks are given in Sec.~\ref{concl}.

\section{Spin dynamics}
\label{theory}

We address the streaming regime accounting for the spin-orbit interaction and elastic scattering by impurities.
We consider a semiconductor A$_3$B$_5$ heterostructure grown along the axis ${z\parallel[001]}$ and choose the in-plane axes as $x\parallel[1\bar10]$ and ${y\parallel[110]}$. In this coordinate frame, the linear in momentum Hamiltonian of spin-orbit interaction takes the form~\cite{AverkievGolub}
\begin{equation}
 {\cal H}_{SO}=\beta_{yx}\sigma_yp_x+\beta_{xy}\sigma_xp_y=\frac{\hbar}{2}\bm \sigma\cdot\bm\Omega(\bm p),
 \label{HSO}
\end{equation} 
where $\bm p$ is the two-dimensional electron momentum, $\sigma_i$ ($i=x,y$) are the Pauli matrices,  $\beta_{xy},\beta_{yx}$ are the spin splitting constants, and ${\bm \Omega}({\bm p})$ is the effective frequency of spin precession caused by Rashba and  Dresselhaus effects. 

Electron spin dynamics is described by a kinetic theory.
The kinetic equation for the spin distribution function ${\bm S}_{\bm p}(t)$ in the passive region reads~\cite{Golub2013,Golub2014}:
\begin{multline}
\left(\frac{\partial}{\partial t}+\frac{1}{\tau_p}+e{\cal E}\frac{\partial}{\partial p_x} \right) {\bm S}_{\bm p}(t)\\=\int\d^2{p}' W_{\bm{pp}'} {\bm S}_{{\bm p}'}(t)+{\bm \Omega}(\bm p)\times{\bm S}_{\bm p}(t).
\label{mainKinetic}
\end{multline} 
Here $W_{\bm p\bm p'}=\delta(p^2-p'^2)/(\pi\tau_p)$ is the probability of elastic scattering from ${\bm p'}$ to $\bm p$ with $\tau_p$ being the elastic scattering time, and we assume the electric field to be applied opposite to the $x\parallel[1\bar10]$ axis. 
After each scattering, an electron accelerates in the $x>0$ direction until reaching the border of the passive region, and then returns to $\bm p=0$. Such a trajectory is shown by arrows in Fig.~\ref{fig:illustration}.

Spin relaxation in the streaming regime  can be caused by both electron penetration into the active region (${p>p_0}$) and by elastic scattering. The latter mechanism is shown to be somewhat more efficient~\cite{Golub2014}, therefore, for the sake of simplicity, we will neglect the first mechanism in this study. To that end we assume that the optical phonon emission time is infinitely short, so the spin distribution is nonzero only in the passive region ($p<p_0$). Accordingly, electrons have zero energy and zero momentum immediately after optical phonon emission. Hence we can separate the two contributions to the spin distribution function
\begin{equation}
 {\bm S}_{\bm p}(t)=\delta(p_y)\theta(p_x){\bm S}_{p_x}^n(t)+{\bm S}^{out}(\bm p;t).
\end{equation} 
The first contribution describes the spin density in the needle ($p_y=0$), while the second one stands for the spin distribution in all the passive region out of the needle.

The streaming regime can be realized only if the elastic scattering is weak ($\tau_p \gg \ttr$), therefore we will consider spin dynamics up to the first order in the small parameter $\ttr /\tau_p \ll  1$. Moreover, we are aimed to solve the problem at the timescale $\sim\tau_p$ or longer, therefore we assume that  the majority of the carriers are in the needle at the time $t=0$. Anyway this situation always establishes during the time $\sim 2\ttr$
after the electric field is switched on. Hence the kinetic equations for the two components of the spin distribution read
\begin{subequations}
\begin{equation}
  \left(\frac{\partial}{\partial t}+\frac{1}{\tau_p}+e{\cal E}\frac{\partial}{\partial p_x} \right) {\bm S}_{p_x}^n(t)={\bm \Omega}(\bm p)\times{\bm S}_{p_x}^n(t),
  \label{prec_needle}
\end{equation} 
\begin{equation}
   \left(\frac{\partial}{\partial t}+e{\cal E}\frac{\partial}{\partial p_x} \right) {\bm S}^{out}(\bm p;t)={\bm \Omega}({\bm p})\times{\bm S}^{out}(\bm p;t)+ \frac{{\bm S}_{p}^n(t)}{2\pi p \tau_p}.
   \label{prec_out}
\end{equation} 
\label{prec}
\end{subequations}

Since the electron trajectories are closed in the $\bm p$-space, the general solution of Eq.~\eqref{prec_needle} can be presented as a superposition of discrete spin modes:
\begin{multline}
  {{\bm S}_{p_x}^n(t)}=\sum_{n=-\infty}^\infty\sum_{l=-1}^1 \frac{g_n^{(l)}}{p_0} \hat{\cal R}(p_x) {\bm e}_n^{(l)}\\
  \times\exp\left[{-\i\omega_n^{(l)}(t-p_x\ttr/p_0)-p_x\ttr/(p_0\tau_p)}\right].
  \label{needle_general}
\end{multline} 
Here $n$ enumerates the modes, $l=-1,0,1$ distinguishes different orientations of the normalized eigenvectors ${\bm e}_n^{(l)}$, $\omega_n$ are complex eigenfrequencies of the system, $g_n^{(l)}$ are the coefficients, and $\hat{\cal R}(p_x)$ is the operator of rotation around the $y$ axis by the angle
\begin{equation}
\label{Phi}
  \Phi(p_x^2) = \int\limits_0^{p_x} {\d p_x \over e{\cal E}} \Omega_y(p_x) = \beta_{yx}\ttr p_x^2/(\hbar p_0).
\end{equation}
The appearance of the operator $\hat{\cal R}(p_x)$ is caused by the fact that inside the needle the precession frequency $\Omega_y(p_x)$ is nonzero, and the electron spins are rotated in the $(zx)$~plane.
Once the spin distribution in the needle is known at $t=0$, the coefficients $g_n^{(l)}$ can be calculated as
\begin{equation}
 g_n^{(l)}=\int\limits_{0}^{p_0}\d p_x\e^{-\i\omega_n^{(l)}\ttr p_x/p_0}{{\bm e}_n^{(l)}}^*\hat{\cal R}^{-1}(p_x)\bm S_{p_x}^n(0),
 \label{gn}
\end{equation}
where we have omitted the terms proportional to the first and higher powers of $\ttr/\tau_p$.
The coefficients $g_n^{(l)}$ depend on excitation conditions: At resonant spin excitation in the vicinity of $\bm p = 0$, $g_n^{(l)}$ are of the same order for all $n$, while at non-resonant excitation the spin distribution at $t=0$ is a smooth function of $\bm p$, and $g_n^{(l)}$  drop with $n$ as ${\propto 1/n^2}$.

The spin distribution outside the needle, ${\bm S}^{out}(\bm p;t)$, can be readily found by integration of Eq.~\eqref{prec_out}:
\begin{equation}
 {\bm S}^{out}(\bm p;t)= \int\limits_{-p_{0x}}^{p_{0x}} \d p_x' \hat{\cal G}_{\bm p\bm p'}  { {\bm S}_{p'}^n(t-{\ttr}(p_x-p_x')/p_0) \over 2\pi \tau_p \sqrt{p_x^{\prime 2} + p_y^2}}.
 \label{out_general}
\end{equation} 
Here 
\[
  p_{0x}=\sqrt{p_0^2-p_y^2},
\]
and the tensor operator $\hat{\cal G}_{\bm p\bm p'}$ denotes Green function of the ordinary differential equation
\begin{equation}
 e{\cal E}\frac{\partial}{\partial p_x}{\cal G}^{\alpha\beta}_{\bm p\bm p'}-\epsilon_{\alpha\gamma\delta}\Omega_\gamma({\bm p}){\cal G}^{\delta\beta}_{\bm p\bm p'}=\delta(p_x-p_x')\delta_{\alpha\beta},
 \label{Green}
\end{equation} 
where the Greek subscripts and superscripts denote the Cartesian components, and $\epsilon_{\alpha\beta\gamma}$ is the Levi-Civita symbol.
The electrons are immediately taken off the left semicircle in Fig.~\ref{fig:illustration} by the electric field, thus
\begin{equation}
\nonumber
{\bm S}^{out}(-p_{0x},p_y;t)=0.
\end{equation} 
In order to satisfy this boundary condition we take the Green function, $\hat{\cal G}_{\bm p\bm p'}$, to be zero at $p_x<p_x'$.

In order to find the eigenfrequencies and spin modes in the streaming regime, one has to substitute general Eqs.~\eqref{needle_general} and~\eqref{out_general} into the boundary condition
\begin{equation}
 {\bm S}_{0}^n(t)-{\bm S}_{p_0}^n(t)=\int\limits_{-p_0}^{p_0} \d p_y {\bm S}^{out}(p_{0x},p_y;t).
 \label{boundary}
\end{equation} 
This condition reflects the fact of immediate optical phonon emission at reaching the border of the active region (the right semicircle in Fig.~\ref{fig:illustration}). It is violated if the electron-phonon scattering is spin-dependent, but this effect is estimated to be negligible in the streaming regime for typical structure parameters~\cite{Golub2013}.
From the coupled set of Eqs.~\eqref{out_general} and~\eqref{boundary} we obtain:
\begin{equation}
{\bm S}_{0}^n(t)-{\bm S}_{p_0}^n(t)={1\over 2\pi \tau_p} \int {\d^2p \over p} {\hat{\cal G}_{{\bm p}_0 \bm p}} {\bm S}^n_p(t-\ttr(p_{0x}-p_x)/p_0),
\label{closed}
\end{equation}
where ${\bm p}_0 = (p_{0x},p_y)$. 

We solve this equation using the perturbation theory in the small parameter $\ttr/\tau_p$. In the absence of elastic scattering ($\ttr/\tau_p =0$), 
it follows from Eq.~\eqref{needle_general} 
that the eigenfrequencies $\tilde\omega_n^{(l)}$ and eigenvectors $\tilde{\bm e}_n^{(l)}$ satisfy the equation
\begin{equation}
   \hat{\cal R}(p_0) \tilde{\bm e}_n^{(l)}=\e^{-\i\tilde\omega_n^{(l)}\ttr}\tilde{\bm e}_n^{(l)}.
\label{eigenvectors}
\end{equation} 
Here $\hat{\cal R}(p_0)$ is an operator of the electron spin rotation after the travel through all the needle. The corresponding rotation angle, see Eq.~\eqref{Phi}, is $\Phi(p_0^2)=\odr\ttr$ with
\begin{equation}
	\odr=\beta_{yx}p_0/\hbar
\end{equation}
being the average precession frequency in the needle, hereafter we assume that $\beta_{yx}\ge0$.
As a result, we find that the eigenfrequencies are combinations of multiples of the travel and drift frequencies: 
\[
\tilde\omega_n^{(l)}=l\left(2\pi n/\ttr-\odr\right), 
\]
and 
$\tilde{\bm e}_n^{(0)} =\hat{\bm y}$, $\tilde{\bm e}_n^{(\pm 1)} = (\hat{\bm z} \pm \i \hat{\bm x})/\sqrt{2}$ with $\hat{\bm x},\hat{\bm y},\hat{\bm z}$ being unit vectors along the Cartesian axes. Physically these frequencies reflect the fact that $S_y^{n}(p_x)$ does not precess, but oscillates in time with the period $\ttr$. By contrast, the spin polarization in $(zx)$ plane precess with the frequency $\odr$ in addition to the periodic oscillations.

In the first order of the perturbation theory the eigenfrequencies become complex and can be presented as ${ \omega_n^{(l)}=\tilde\omega_n^{(l)}+\delta_n^{(l)} }$~\cite{Levinson}. Provided
\begin{equation}
 \odr\tau_p\gg1
 \label{degenerate}
\end{equation} 
to account for the elastic scattering, we substitute the eigenvectors $\tilde{\bm e}_n^{(l)}$ into Eq.~\eqref{needle_general} find $\bm S^n_{p_x}$, and then 
Eq.~\eqref{closed} yields the corrections $\delta_n^{(l)}$. In the opposite case ${\odr\tau_p\lesssim 1}$, the perturbation theory for degenerate levels should be used.

The spin dynamics can be probed by Faraday, Kerr or ellipticity signals which are determined by the total electron spin, $\bm S(t)$. In the streaming regime most of the particles are localized inside the needle, therefore we have
\begin{equation}
 \bm S(t)=\int\limits_0^{p_0}\d p_x \bm S_{p_x}^n(t) = \sum_{n=-\infty}^\infty\sum_{l=-1}^1 g_n^{(l)} \bm s_n^{(l)}\e^{-\i\omega_n^{(l)}t}.
\label{total_spin}
\end{equation}
Here we have introduced the average spin polarization in the $n$th mode
\begin{equation}
 \bm s_n^{(l)}=\int\limits_0^{p_0}\frac{\d p_x}{p_0} \hat{\cal R}(p_x) \tilde{{\bm e}}_n^{(l)}\e^{\i\tilde{\omega}_n^{(l)}\ttr p_x/p_0}.
\label{sn}
\end{equation} 

The results of this Section describe spin dynamics at arbitrary strong and anisotropic spin-orbit splitting. 

\section{Results and Discussion}
\label{results}

In general case the Green function $\hat{\cal G}_{\bm p\bm p'}$ can not be found analytically, therefore in the next subsections we consider separately two limits: (i) the effective field ${\bm \Omega}(\bm p)$ is oriented along $x$-axis ($\beta_{yx}=0, \beta_{xy}\neq0$) and (ii) ${\bm \Omega}(\bm p) \parallel y$ ($\beta_{xy}=0, \beta_{yx}\neq 0$).
In the end of this Section, we briefly analyze spin dynamics in the presence of both components in ${\bm \Omega}(\bm p)$.

\subsection{Longitudinal effective field \texorpdfstring{$\bm{\Omega}(\bm p)\parallel x$}{Omega_x}}
\label{sec_omegaX}

First we consider the limit of $\beta_{yx}=0$, when the effective field $\bm\Omega(\bm p)$ is parallel to the electric field. Since $\Omega_x \propto p_y$, the spin precession in the needle is absent, and  Eq.~\eqref{eigenvectors} is simplified to $\exp(-\i\tilde\omega_n^{(l)}\ttr)=1$, which yields
\begin{equation}
 \tilde\omega_n^{(l)}=2\pi n/\ttr.
 \label{omegas_xy}
\end{equation} 
Clearly in this limit the condition Eq.~\eqref{degenerate} fails, and below we apply the perturbation theory for degenerate levels.

Out of the needle, electrons move ballistically conserving the $p_y$ momentum component. Therefore,
the electron spin rotates in the $(yz)$ plane with the constant frequency $2\beta_{xy}p_y/\hbar$. Hence the Green function of Eq.~\eqref{Green}, $\hat{\cal G}_{\bm{pp}'}$, is the identity operator for $S_{x}$, and it multiplies
{$S_{y}(\bm p;t)\pm\i S_{z}(\bm p;t)$} in Eq.~\eqref{closed}
by the factors $\exp{[\pm {\rm i}\Omega_x(p_x-p_x')/(e\mathcal{E})]}$. 
On average, scattered electrons have zero momentum $p_y$, therefore the spin polarization does not precess even with account for scattering off the needle. Hence we can choose the basis vectors as ${\bm{e}_n^{(l)}=\hat{\bm y},\hat{\bm x}, \hat{\bm z} }$ for ${l=-1,0,1}$, respectively. 
Substituting these $\bm{e}_n^{(l)}$ and $\tilde\omega_n^{(l)}$ from Eq.~\eqref{omegas_xy} into Eqs.~\eqref{needle_general} and~\eqref{closed}, we obtain the corrections to the eigenfrequencies in the form
\begin{multline}
 \delta_n^{(l)}=-\frac{\i}{\tau_p}+\frac{\i}{2\pi\tau_p}\int\frac{\d^2{p}}{pp_0}\exp\left[2\pi\i n\left(p_{0x}-p_x+p\right)/p_0\right]\\
 \times\cos\left[2l\beta_{xy}p_y(p_{0x}-p_x)\ttr/(\hbar p_0)\right].
 \label{deltas_xy}
\end{multline}
Hereafter we assume that all integrations are performed over the passive region $p<p_0$ only.

\begin{figure}[t]
\includegraphics[width=\linewidth]{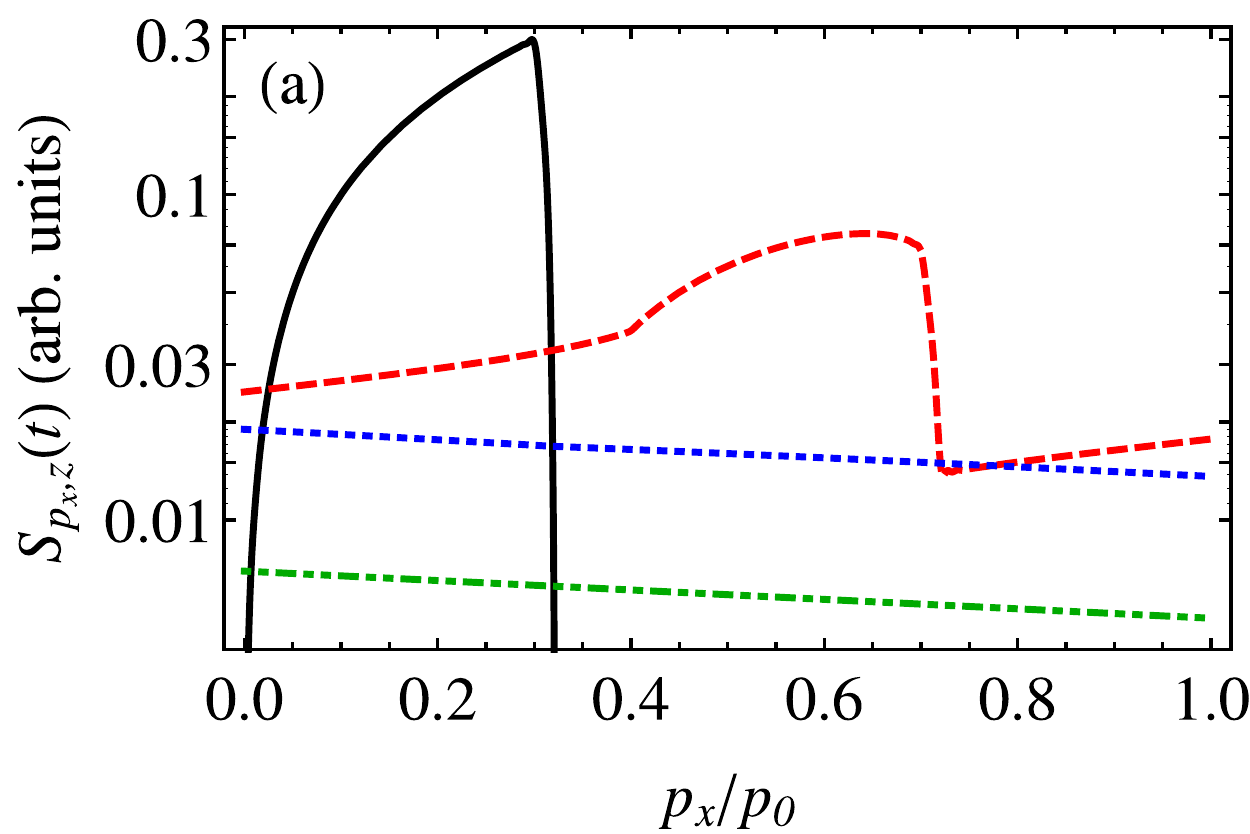}
\includegraphics[width=\linewidth]{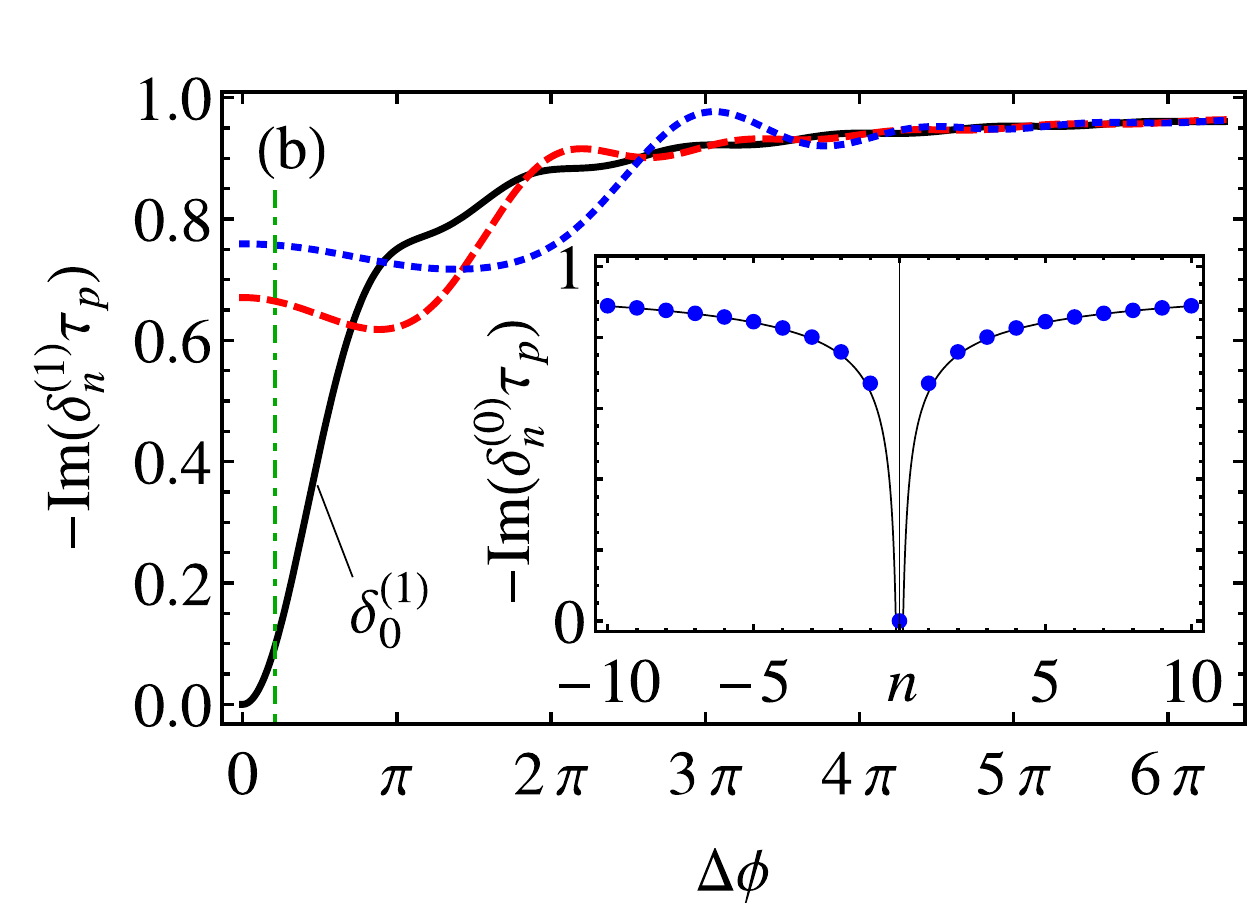}
\caption{(a) The spin distribution {in the logarithmic scale} for $\tau_p={3\ttr}$, $\tau_s^z=10\tau_p$ at the times $t=0$ (black solid curve), $1.8\tau_p$ (red dashed curve), $10\tau_p$ (blue dotted curve) and $20\tau_p$ (green dash-dotted curve).
(b) The decay rates of the first three $z$-spin modes for $n=0$ (black solid curve), $n=1$ (red dashed curve), and $n=2$ (blue dotted curve) as functions of $\Delta \phi$, Eq.~\eqref{dphi}.
The green dash-dotted line denote the spin splitting corresponding to the panel (a).
The inset shows the decay rates of the $x$-modes
(dots) and the analytical approximation Eq.~\eqref{deltas_anal} (solid line).
}
\label{fig:density_relax}
\end{figure}

Despite the developed formalism is quite cumbersome, its interpretation is straightforward.
Let us consider the dynamics of $x$ spin component. It corresponds to $l=0$, and for a homogeneous distribution ($n=0$) one finds $\delta_0^{(0)}=0$, i.e. the eigenfrequency is zero.
This reflects the fact that the total spin polarization along the $x$ axis is conserved, because it is not affected by the spin-orbit interaction. Therefore its dynamics is the same as for the particle distribution function. The decay rates of the excited $x$-modes [Eq.~\eqref{deltas_xy} for $l=0$ and $n\neq0$] are nonzero. 
They are presented in the inset to Fig.~\ref{fig:density_relax}(b) which shows that all of them are of the same order ($\sim 1/\tau_p$). 
This means that the spin distribution relaxes during the time $\tau_p$ to a constant (zero mode). In the limit of $|n| \gg 1$ one can find from Eq.~\eqref{deltas_xy}:
\begin{equation}
 -\Im\delta_n^{(0)}\approx\frac{1}{\tau_p}\left(1-\frac{1}{\sqrt{8|n|}}\right).
 \label{deltas_anal}
\end{equation}
This analytical expression describes the decay of all modes except for $n = 0$ with accuracy of $4$~\%, see the inset to Fig.~\ref{fig:density_relax}(b).

The spin components $S_{\bm p,z}^{out}$ and $S_{\bm p,y}^{out}$ precess, which results in spin relaxation. First we consider the limit 
where characteristic spin rotation angles out of the needle
\begin{equation}
\label{dphi}
 \Delta\phi=\beta_{xy}p_0\ttr/\hbar
\end{equation} 
are small: $\Delta\phi\ll1$.
In this case all the excited modes relax with approximately the same rates as the $x$-modes, i.e. ${\delta_n^{(\pm 1)}\approx\delta_n^{(0)} \sim 1/\tau_p}$ for $n\neq0$. The decay rates of the homogeneous spin distribution $1/\tau_s^{z,y}=-\Im\delta_0^{(\pm1)}$ are given by~\cite{fnote}
\begin{equation}
 \frac{1}{\tau_s^{z}}=\frac{1}{\tau_s^{y}}={7\over 30} {\left( \Delta\phi\right)^2  \over \tau_p}.
\label{tau_s_zy}
\end{equation} 
One can see that the spin relaxation time in this limit is much longer than $\tau_p$. This means that the spin distribution relaxes in two stages. First, after a time $\sim\tau_p$ all the spin modes except for $n=0$ decay, so the spin distribution in the needle becomes nearly uniform. In the second stage, this uniform spin distribution decays with the rate $1/\tau_s^{z}$.
This two-stage relaxation is illustrated in Fig.~\ref{fig:density_relax}(a).
Interestingly, since the decay rates of all the excited modes are of the same order, the spin distribution almost conserves its shape during the relaxation, so that after the time $\sim 2\tau_p$ the initial shape of the distribution is still visible.

The advantage of the presented theory is that it can describe spin dynamics and relaxation for arbitrary characteristic rotation angle {$\Delta\phi$}. If this parameter is of the order of unity, then all the spin modes decay with the rate $\sim1/\tau_p$. The decay rates of the first three modes calculated by Eq.~\eqref{deltas_xy} are presented in Fig.~\ref{fig:density_relax}(b). In the limit of ${\Delta\phi}\to\infty$,
$\delta_n=-\i/\tau_p$ for any $n$, because a single elastic scattering is sufficient for complete spin dephasing.

One can see that for the simple eigenfrequencies Eq.~\eqref{omegas_xy}, the average spin polarization is nonzero only in the zeroth mode, since only the term with $n=0$ contributes to $S_z(t)$, see Eq.~\eqref{sn}. Therefore {the total spin decays monoexponentially with the}
rate $-\Im\delta_0^{(1)}$ 
for arbitrary values of $\beta_{xy}$ despite in this case all the spin modes decay at different timescales. This spin relaxation rate is shown by the black solid curve in Fig.~\ref{fig:density_relax}(b) as a function of the spin-orbit coupling strength.
Note that the decay rates of all the modes are always smaller than $\tau_p^{-1}$ because at least one scattering is needed to randomize spin orientation.
The series of local maxima in the spin relaxation rate presented in Fig.~\ref{fig:density_relax}(b) roughly corresponds to the condition $\Delta\phi=n\pi$. Qualitatively, these maxima and the oscillations in the decay rates of the other modes, are related to the destructive interference of the spin rotation angles for scattered electrons.

\subsection{Transverse effective field \texorpdfstring{$\bm{\Omega}(\bm p)\parallel y$}{Omega_y}}
\label{transverse}

If the effective field is perpendicular to the electric field, the dynamics of $y$ spin component is the same as for the spin-independent distribution function.
By contrast, the spin components lying in the $(zx)$ plane feel the effective field $\Omega_y \propto p_x$, so the characteristic precession angles are not small even inside the needle where $p_x \leq p_0$. Therefore we concentrate on these components.

At $\beta_{xy}=0$, one can use the unperturbed basis vectors and eigenfrequencies $\tilde {\bm e}_n^{(l)}$ and $\tilde\omega_n^{(l)}$.
We find the Green function of Eq.~\eqref{Green} in the form $\hat{\cal G}_{\bm{pp}'}=\hat{\cal R}(p_x)\hat{\cal R}^{-1}(p_x')$, and from Eq.~\eqref{closed} we obtain
\begin{align}
\label{deltas_yx}
  \delta_n^{(l)}= & -\frac{\i}{\tau_p}  + \i\frac{\e^{\i l\odr\ttr}}{2\pi\tau_p} \\
&	\times \int\frac{\d^2{p}}{p_0p}
   \exp\left[\i l\left(2\pi n-\odr\ttr\right)\left(p_{0x}-p_x+p\right)/p_0\right]. \nonumber
\end{align} 
One can see that real parts of $\delta_n^{(l)}$ have opposite signs for $l=\pm 1$, while their imaginary parts coincide.
Importantly, it follows from Eq.~\eqref{deltas_yx} that for all $l$ ${\delta_n^{(l)}(\odr)=\delta_0^{(l)}(\odr-2\pi n/\ttr)}$
and, in particular, 
\[ \delta_n^{(l)}(2\pi n/\ttr)=0.
\] 
This means that the $n$th spin mode does not decay at $\odr=2\pi n/\ttr$. On the other hand, it follows from Eq.~\eqref{needle_general} that, at $\odr\neq 0$, the average spin polarization is nonzero in any mode.  
Therefore the decay time for the spin lying in the $(zx)$ plane can be infinitely long.

\begin{figure}[t]
\includegraphics[width=\linewidth]{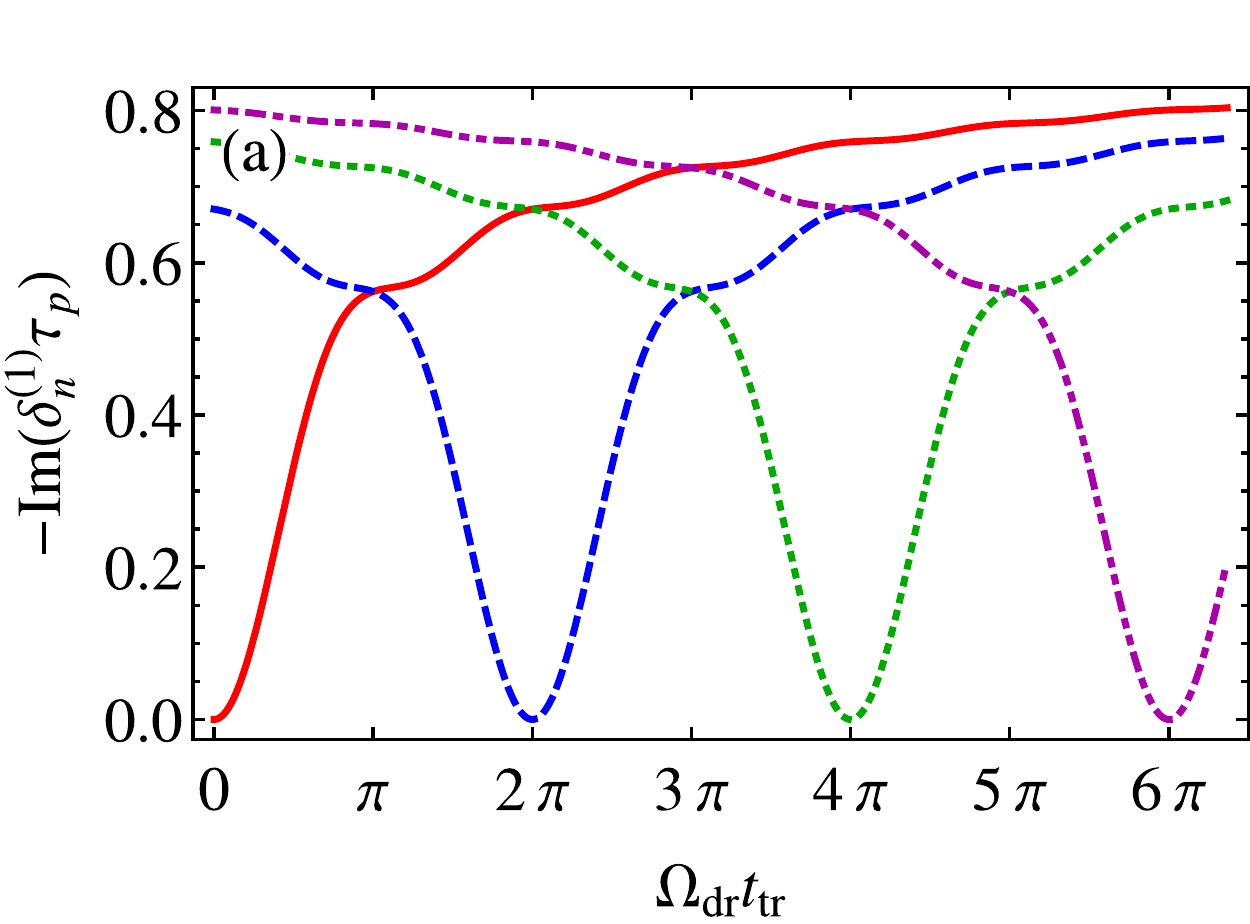}
\includegraphics[width=\linewidth]{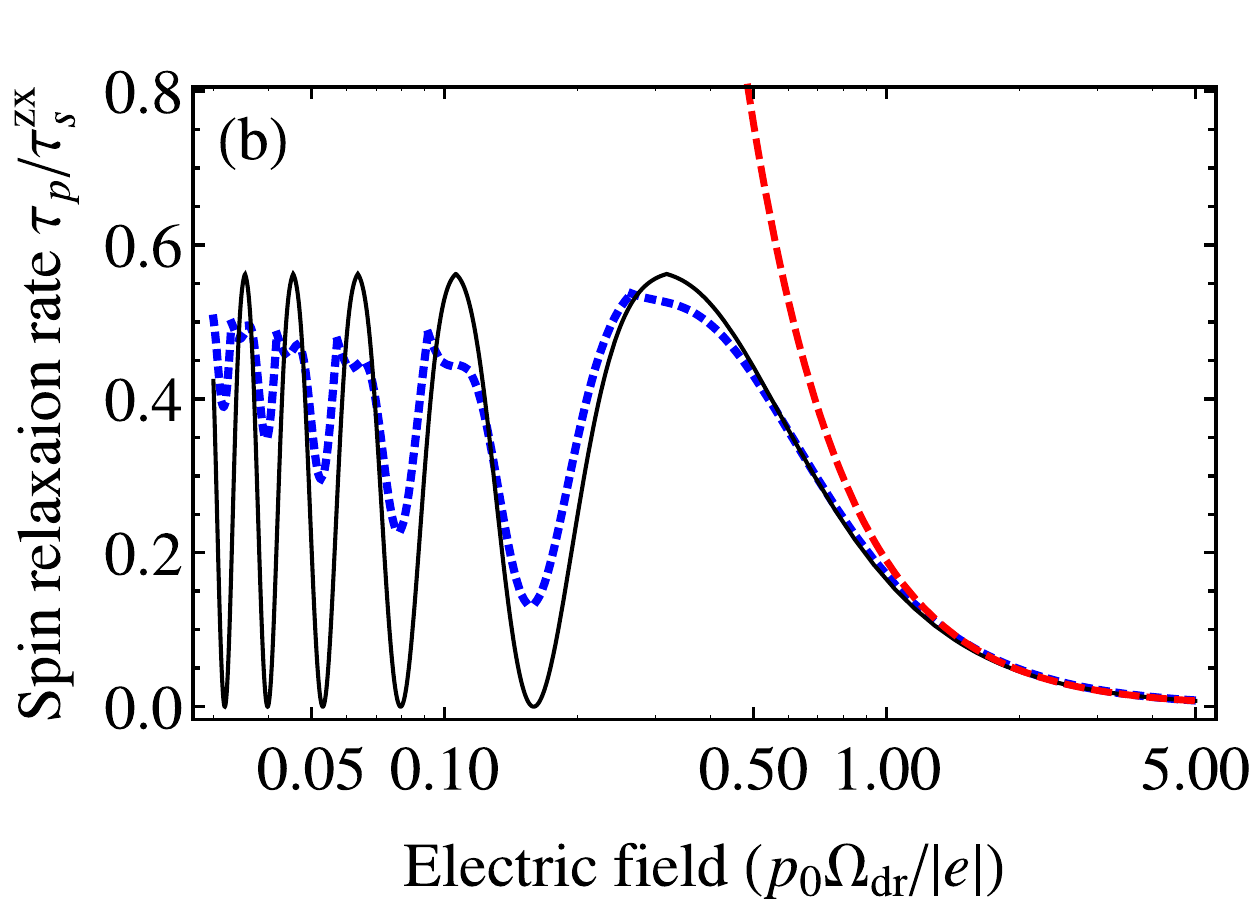}
\caption{
(a) The decay rates of the first four spin modes, $-\Im(\delta_n^{(1)}\tau_p)$ 
for $n=0$ (red solid curve), $n=1$ (blue dashed curve), $n=2$ (green dotted curve) and $n=3$ (magenta dash-dotted curve). 
(b) Spin relaxation rate of the most long-living mode
as a function of electric field (black solid curve) and its approximation Eq.~\eqref{tau_yx} (red dashed curve). The spin relaxation rate in the presence of both components of the effective field is shown by the blue dotted curve at ${\beta_{xy}=\beta_{yx}/3}$.
}
\label{fig:spin_relax}
\end{figure}

The fact that the spin relaxation time is infinite for $\odr\ttr=2\pi n$ can be explained by considering an electron's motion from $p_x$ to $p_x'$ in the ballistic area. Its spin is rotated by the angle ${\Phi({p_x'}^2)-\Phi(p_x^2)}$ around the $y$-axis, where $\Phi\sim p_x^2$ is given by Eq.~\eqref{Phi}.
Since $p_y$ momentum component is constant at this motion, the rotation angle can be rewritten as
${\Phi({p'}^2)-\Phi(p^2)}$.
Moreover, during spin-independent elastic  scattering, $p^2$ is conserved, and the spin direction is not changed. Therefore, if $\bm s$ is a spin of a given electron, then the quantity
\begin{equation}
 {\cal I}=(s_z+\i s_x)\e^{-\i\odr\ttr p^2/p_0^2}
\label{inv}
\end{equation} 
is conserved during an electron's motion in the passive region, and this is correct in all orders in $\ttr/\tau_p$.
Here we assume, as before, that the optical phonon emission time is infinitely short.
Equation~\eqref{inv} demonstrates that when an electron reaches the active region, its spin is rotated around the $y$ axis by the angle $\odr\ttr$ 
irrespectively to a number of elastic scatterings. Therefore at $\odr\ttr=2\pi n$ the electron spin always returns to its initial direction after an optical phonon emission, i.e. spin relaxation is absent.
This situation is similar to the persistent spin helix~\cite{Bernevig06}, but here it is realized in energy domain. The analogy comes from the fixed relation between the electron displacement along the field direction, $\Delta x$, and the gain of its energy $\Delta E=|e {\cal E}| \Delta x$ independent of the electron trajectory. 

In general case the spin relaxation time, $\tau_s^{zx}$, corresponds to the smallest decay rate of all the modes:
\begin{equation}
 1/\tau_s^{zx} =\min(-\Im\delta_n^{(1)})\equiv-\Im\delta_{n^*}^{(1)},
\label{tau_s_zx}
\end{equation}
where
\begin{equation}
{n^*=[\odr\ttr/(2\pi)+1/2]}
\label{n_star}
\end{equation}
is the number of the most long living mode with $[x]$ being the integer part of $x$. 
The decay rates of the spin modes calculated after Eq.~\eqref{deltas_yx} are presented in Fig.~\ref{fig:spin_relax}(a).
From  this figure one can see 
that the spin relaxation time defined {by Eq.~\eqref{tau_s_zx} } is a periodic function of $\odr\ttr$. Despite $\odr$ is the structure parameter, the travel time $\ttr$ can be changed by the electric field. Figure~\ref{fig:spin_relax}(b) illustrates the oscillations of the spin relaxation rate as a function of the applied electric field in the streaming regime. 
In the limit $\odr\ttr\ll 1$ we have from Eq.~\eqref{deltas_yx}:
\begin{equation}
 \delta_0^{(\pm 1)} \approx \pm \frac{\odr\ttr}{2\pi\tau_p}\left(2+4G-\pi\right)-\i\frac{(\Omega_{\rm{dr}} t_{\rm{tr}})^2}{3\pi\tau_p}\left(2\pi+1-6G\right),
 \label{tau_yx}
\end{equation} 
where $G\approx0.9159\ldots$ is the Catalan constant. This asymptotic for the spin relaxation time is plotted in Fig.~\ref{fig:spin_relax}(b) by a red dashed curve. Comparison with the numerical calculation shows that this expression is correct up to $\odr\ttr \sim 1$.

\begin{figure}[t]
\includegraphics[width=\linewidth]{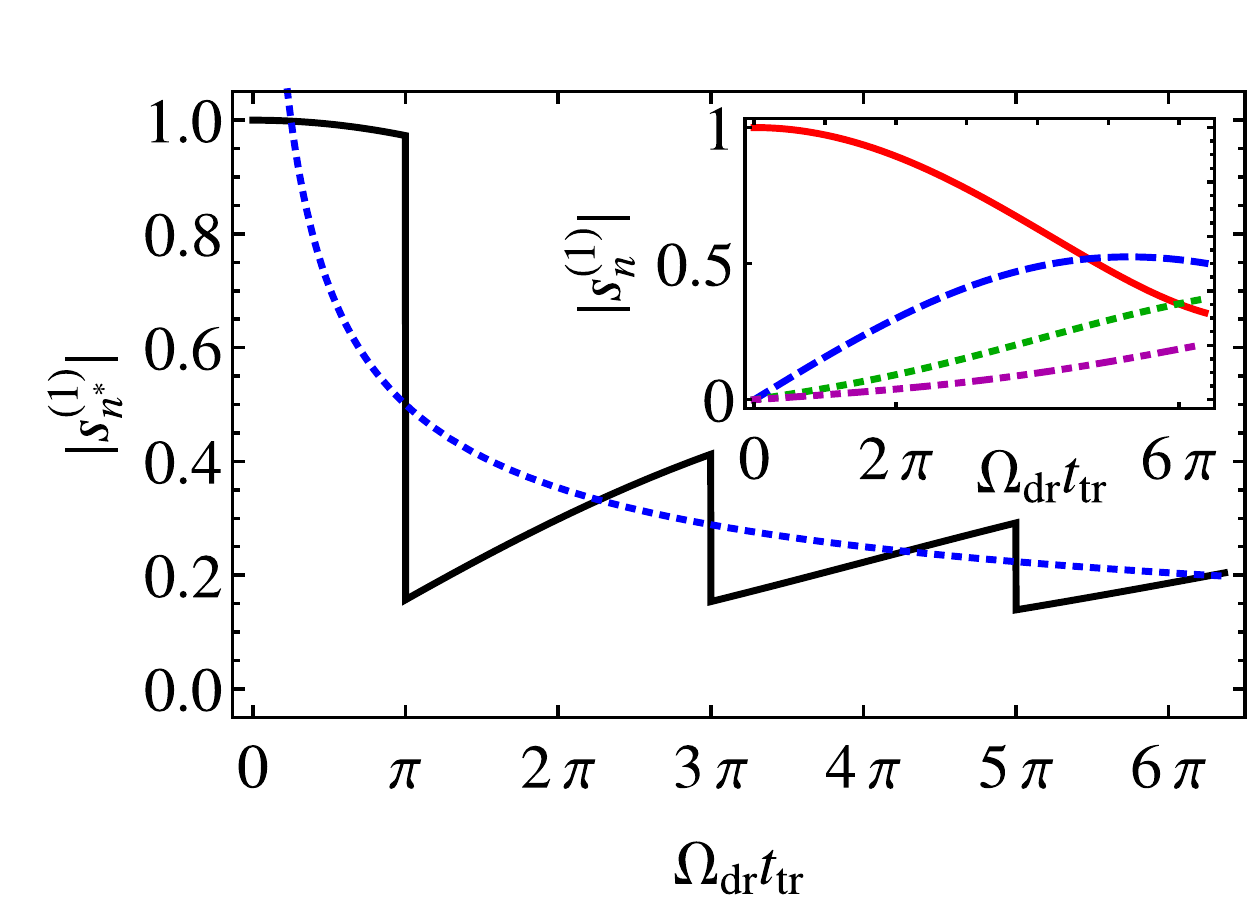}
\caption{
Average spin $s_{n^*}$ in the most long-living mode 
as a function of $\odr\ttr$ (black solid curve) and its asymptotic, Eq.~\eqref{sn_star} (blue dotted curve). Inset: Average spin in the $n$th mode, $s_n$, 
for $n=0$ (red solid curve), $1$ (blue dashed curve), $2$ (green dotted curve) and $3$ (magenta dash-dotted curve).
}
\label{fig:modes}
\end{figure}

After decay of all the modes except for the $n^*$th one, the total spin polarization is defined by two vector coefficients, $\bm s_{n^*}^{(\pm 1)}$, see Eq.~\eqref{total_spin}, and it lies in the $(zx)$ plane. 
Their absolute values coincide  and decrease as a function of $n^*$.
The inset to Fig.~\ref{fig:modes} shows the quantity
\[
\left|s_{n}^{(1)}\right| = \sqrt{\left|s_{n,z}^{(1)}\right|^2 + \left|s_{n,x}^{(1)}\right|^2}
\] 
as a function of the parameter $\odr\ttr$ for the first four modes. For $\odr\ttr=0$,  average spin is present only in the zeroth mode.
With increase of $\odr\ttr$, the non-zero spin polarization arises in all the other modes due to a dependence of the spin precession frequency on $p_x$.
Figure~\ref{fig:modes} shows $|s_{n^*}^{(1)}|$
as a function of $\odr\ttr$. One can see that the average spin polarization in the most long living mode exhibits damped oscillations. The jumps in the dependence $|s_{n^*}^{(1)}|$ are related to the switching between the modes $n$ and $n+1$ at the points $\odr\ttr=2\pi(n+1/2)$, see Eq.~\eqref{n_star}. For large $\odr\ttr$ the conserved spin decays as
\begin{equation}
 |s_{n^*}^{(1)}|\approx\frac{1}{2}\sqrt{\frac{\pi}{\odr\ttr}}.
\label{sn_star}
\end{equation} 
This dependence is plotted in Fig.~\ref{fig:modes} by a dotted line.

If both components, $\Omega_x$ and $\Omega_y$ are nonzero, a numerical solution of Eq.~\eqref{Green} is needed in order to describe the spin dynamics. This situation is qualitatively similar to the case of electric field orientation at arbitrary angle to the main axes. The results of calculations for $\beta_{xy}\neq 0$ are shown in Fig.~\ref{fig:spin_relax}(b) by a blue  dotted line. This dependence demonstrates that the $n$th mode decays even at $\odr\ttr=2\pi n$ if $\Omega_x$ is nonzero. Spin relaxation is switched on due to an additional phase $\propto \beta_{xy}$ acquired by electrons which violates invariance of ${\cal I}$, Eq.~\eqref{inv}.
However, pronounced drops in the decay rate at resonant conditions $\odr\ttr=2\pi n$ are still present.

\section{Spin noise}
\label{noise}

The analysis of the spin dynamics performed in the previous Sections reveals multiple timescales in the streaming regime:
\[
\ttr\ll\tau_p<\tau_s.
\]
Therefore studies in the frequency domain are useful, and the spin noise spectroscopy serves as an excellent tool to that end. 
In the steady state the majority of the electrons are in the needle, so the spin fluctuations are characterized by the correlation functions
\begin{equation}
 \left\langle \delta S_{p_x,\alpha}^n(t) \delta S_{p_x',\beta}^n(t+\tau)\right\rangle,
\end{equation} 
where angular brackets denote averaging over the time $t$ for the given delay $\tau$. 
In the steady state, the distribution function in the needle is uniform, so the one-time correlator is given by~\cite{ll10_eng}
\begin{equation}
 \left\langle \delta S_{p_x,\alpha}^n(t) \delta S_{p_x',\beta}^n(t)\right\rangle=
\frac{N}{4p_0} \delta(p_x-p_x') \delta_{\alpha\beta}.
\label{1time}
\end{equation} 
Here $N$ is the electron two-dimensional density, and we ignore an electric field induced spin polarization because its value does not exceed a few per cents in the streaming regime~\cite{Golub2013,Golub2014}.

Ultimately we are interested in the total spin correlation function, $\left\langle S_z(t)S_z(t+\tau)\right\rangle$, where we assume the probe beam to propagate along $z$ direction. 
Since a correlator of any physical quantity obeys the same linear kinetic equation as the quantity itself~\cite{Bareikis,Kogan,ll10_eng}, the spin correlation function for $\tau>0$ can be presented in the form
\begin{multline}
	  \left\langle \delta S_z(t)\delta S_z(t+\tau)\right\rangle \\
		=	\int\d p_x \int \d p_x' \int \d p_x''\left\langle{\delta S_{p_x,z}^n(t) T^{z\alpha}_{p_x'p_x''}(\tau)\delta S_{p_x'',\alpha}^n(t)}\right\rangle, \nonumber
\end{multline}
where $T^{\alpha\beta}_{\bm{pp}'}(\tau)$ ($\alpha,\beta=x,y,z$) is the Green function of the kinetic equation~\eqref{mainKinetic}. Due to linearity of $\hat{T}_{p_xp_x'}$ and using Eq.~\eqref{1time}, this expression can be recast as
\begin{equation}
	\left\langle \delta S_z(t) \delta S_z(t+\tau)\right\rangle=\sum_{p_xp_x'}T^{zz}_{p_x'p_x}(\tau)\frac{N}{4p_0}\equiv 
	{1\over N} S_0S_z(\tau).
	\label{time_corr}
\end{equation}
Here $S_0={N}/2$, and  $S_z(\tau)$ is given by Eq.~\eqref{total_spin}, where the spin distribution $S_{p_x,z}^n(\tau)$ is found using the initial condition
\begin{equation}
 S_{p_x,z}^n(0)=S_0/p_0.
\label{init_cond}
\end{equation} 
Accordingly the coefficients $g_n^{(l)}$ can be calculated after Eq.~\eqref{gn}: ${g_n^{(l)}=S_0s_{n,z}^{(l)*}}$, where $\bm s_n^{(l)}$ is given by Eq.~\eqref{sn}.

The spin noise spectrum is defined by
\begin{equation}
 (\delta S_z^2)_\omega=\int\limits_{-\infty}^\infty \left\langle S_z(t)S_z(t+\tau)\right\rangle\e^{\i\omega \tau}\d\tau.
\label{spectr_def}
\end{equation} 
Since the correlator is an even function of $\tau$~\cite{Bareikis,Kogan,ll10_eng}, we obtain from Eqs.~\eqref{time_corr} and~\eqref{spectr_def}:
\begin{equation}
  (\delta S_z^2)_\omega=\frac{N}{4}\sum_{n=-\infty}^\infty\sum_{l=-1}^1
  \left|s_{n,z}^{(l)}\right|^2
	{\rm Im} \left({1\over \omega-\omega_n^{(l)}}+{1\over \omega+\omega_n^{(l)}}\right).
\label{noise_spectrum}
\end{equation} 
This equation demonstrates that the spectrum consists of the series of Lorentzian peaks centered at the eigenfrequencies. The areas of the peaks are proportional to the squared total spin in the corresponding mode, and the widths are determined by the decay rates of the modes.

At $\bm\Omega(\bm p)\parallel x$, the average spin polarization along $z$-axis is nonzero only in the mode characterized by $n=0$ and $l=1$. As it follows from Eq.~\eqref{deltas_xy}, the corresponding eigenfrequency $\omega_0^{(1)}$ is imaginary, and $s_{0,z}^{(1)}=1$, see Eq.~\eqref{sn}. Hence the spin noise spectrum simply reads
\begin{equation}
 (\delta S_z^2)_\omega=\frac{N}{2}\frac{\tau_s^z}{1+(\tau_s^z\omega)^2},
\end{equation} 
where $1/\tau_s^z=-{\rm Im}~\omega_0^{(1)}$ is shown by black solid line in Fig.~\ref{fig:density_relax}(b). At $\Delta\phi \ll 1$, $\tau_s^z$ is given by Eq.~\eqref{tau_s_zy}.

\begin{figure}[t]
\includegraphics[width=\linewidth]{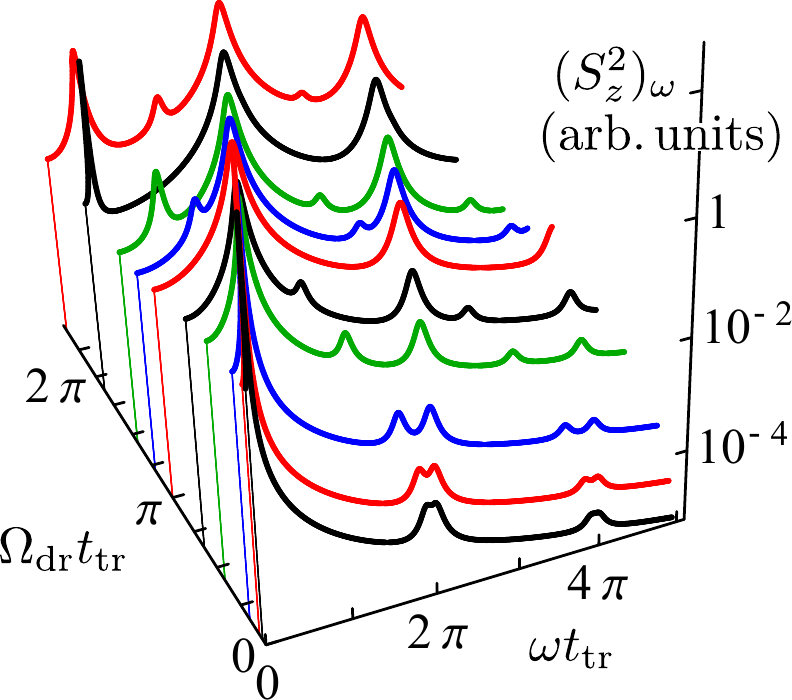}
\caption{Spin noise spectra in the streaming regime calculated after Eq.~\eqref{noise_spectrum} for $\tau_p=3\ttr$ and $\beta_{xy}=0$.}
\label{fig:noise}
\end{figure}

In the opposite case of transverse effective field ${\bm\Omega(\bm p)\parallel y}$, all the spin modes with $l=\pm1$ contribute to the spin noise spectrum. Figure~\ref{fig:noise} demonstrates the spin noise spectra for various values of $\odr\ttr$.
One can see that the spectrum consists of a series of peaks of different widths.

If $\odr\ttr\ll2\pi$,
the dominant contribution to the spin noise spectrum is given by the terms with $n=0$ and $l=\pm 1$. Accordingly, at $\omega>0$ the spin noise spectrum reads
\begin{equation}
  (\delta S_z^2)_\omega \approx \frac{N}{4}\frac{\tau_s^z}{1+[(\omega-\odr){\tau_s^z}]^2},
\end{equation} 
i.e. it has a Lorentzian shape  centered at the frequency $\odr$. The width is $1/\tau_s^z =-\Im \delta_0^{(1)}$, where $\delta_0^{(1)}$ is given by Eq.~\eqref{tau_yx}. 
The other peaks are centered at the frequencies $2\pi n/\ttr\pm\odr$ and have the widths of the order of $1/\tau_p$; their amplitudes decrease as $1/n^4$.
The shift of the  main peak is caused by the nonzero effective precession frequency inside the needle acting as a constant magnetic field. 
According to the fluctuation-dissipation theorem, this shift is
equivalent to the electric-current induced shift of the electron spin resonance spectra~\cite{Jantsch}.

For $\odr\ttr/2\pi \sim 1$,
the widths of all the peaks have an order of $1/\tau_p$, but the amplitudes of the first few peaks are comparable to each other, see Fig.~\ref{fig:noise}. However, when $\odr\ttr$ approaches $2\pi n$,
the spin relaxation time  tends to infinity, see Sec.~\ref{transverse}. As a result, the peak centered at ${\omega=|2\pi n/\ttr-\odr|}$ becomes very high and narrow. 

Let us analyze the situation when the spin-orbit interaction is absent. In this limit
the total spin is conserved, and there are no spin fluctuations. Nevertheless, the spin noise spectroscopy has an access to spin dynamics even in this case. Indeed, in transmission or reflection experiments, the measured spin Faraday or Kerr signals, $\Theta(t)$, are related to the spin distribution function via 
\begin{equation}
	\Theta(t) = \int \d^2p \:  K(p) S_{{\bm p}, z}(t).
	\label{theta_def}
\end{equation}
Here $K(p) \propto (p^2+a^2)^{-1}$, where $a$ is determined by the detuning between the probe beam frequency and the energy gap~\cite{Golub2014}. The resonant dependence $K(p)$ reflects the fact that the electrons with larger energy give smaller contribution to the spin signals, and it 
gives rise to fluctuations of $\Theta(t)$. Analysis shows that the spectrum $\Theta^2_\omega$ defined in analogy with Eq.~\eqref{spectr_def} has the form:
\begin{equation}
  (\delta \Theta^2)_\omega=\frac{N}{4}\sum_{n=-\infty}^\infty\left|\Theta_n\right|^2{\rm Im} \left({1\over \omega-\omega_n^{(0)}}+{1\over \omega+\omega_n^{(0)}}\right),
\label{theta_noise}
\end{equation} 
where 
\begin{equation}
\Theta_n = \int\limits_0^{p_0} {dp \over p_0} K(p) {\e}^{2\pi\i n p/p_0}.
\label{theta_n}
\end{equation}
The eigenfrequencies $\omega_n^{(0)}=2\pi n/\ttr+\delta_n^{(0)}$, where $\delta_n^{(0)}$ should be calculated after Eq.~\eqref{deltas_xy}. 
One can see that the noise spectrum $\Theta^2_\omega$ has a structure of Lorentzian peaks centered at multiples of the travel frequency and having the widths of the order of $1/\tau_p$ (except for $n=0$). 
This spectrum is different from the electric current fluctuation spectrum~\cite{Levinson,Bareikis,Kogan} by the zero frequency peak: since the spin relaxation time is infinite in this limit, the width of the peak is zero.
The above analysis demonstrates that the proposed method extends the spin noise spectroscopy technique to the case when the total spin is conserved and does not fluctuate. The same approach allows measuring, e.g. energy relaxation rate of free electrons if spin relaxation is slow enough.

\section{Conclusions}
\label{concl}

We have developed a kinetic theory of spin dynamics in the streaming regime with account for elastic scattering and spin-orbit interaction. 
The spin eigenmodes are identified and their decay rates are calculated. 
We have shown that electron spin dynamics is strongly different in the limits of small and large spin rotation angle during the time $\ttr$. If it is small, then the spin distribution becomes uniform  inside the needle on the timescale of one elastic scattering event. Afterwards, the average spin polarization monoexponentially decreases with the decay time $\tau_s \gg \tau_p$. In the opposite limit of large rotation angles, the spin relaxation time has an order of $\tau_p$. However, the spin relaxation time oscillates as a function of the electric field and infinitely increases when $\odr\ttr$ approaches a multiple of $2\pi$. 
We have demonstrated that this effect is robust against elastic scattering, and, in fact, it is the energy space analogue of the persistent spin helix.
The pronounced oscillations exist even in the presence of the transverse component of the effective magnetic field.

The spin noise spectrum in the streaming regime is calculated. The spectrum consists of a series of the peaks corresponding to the different spin modes, and the widths of the peaks correspond to the lifetimes of the modes.
The present study demonstrates that the spin noise spectroscopy applied to nonequilibrium electron systems reveals the parameters of spin dynamics and, in particular, 
the spin-orbit splittings.
In the range of low frequencies inherent to the traditional spin noise spectroscopy, evolution of the spectrum reflects the strong oscillations of the spin relaxation rate.
The advantage of the ultrafast spin noise spectroscopy~\cite{Berski-fast-SNS} paves the way for 
observation of peaks in the spin fluctuation spectrum in the streaming regime.
Moreover, if the spin-orbit coupling is small, the resonant measurement of the Faraday or Kerr angle fluctuations allow investigating the particle distribution function dynamics.

As an outlook we note, that spin dynamics in the streaming regime is extremely interesting to investigate in topological insulators. One of the reasons is a high value of the current-induced spin polarization. It is established that the spin polarization in topological insulators is proportional to a ratio of the drift and the Fermi momenta which is small in weak fields~\cite{Golub2011,TI_Japan,Pesin2012,li2014electrical}. By contrast, in the streaming the needle-like electron distribution results in a large drift momentum, and the current-induced spin polarization is 100~\%.

\acknowledgments We thank M.\,M. Glazov for fruitful discussions. Partial  support from 
RFBR and RFBR-DFG ICRC TRR160, Dynasty Foundation, RF President Grant No.~SP-643.2015.5 and Programmes of RAS is gratefully acknowledged.

\end{document}